# A polynomial-based Monte Carlo approach for estimating long-term collision probabilities.

Andrea Zollo[(1)], Cristina Parigini[(2)], Roberto Armellin[(2)], Juan Félix San Juan Díaz[(3)], Annarita Trombetta[(1)] and Ralph Kahle[(1)]

[(1)] *German Space Operations Centre (GSOC), German Aerospace Centre DLR, Münchener Str. 20, 82234 Weßling, Germany, email:*
*{andrea.zollo, annarita.trombetta, ralph.kahle}@dlr.de*
[(2)] *Te Pūnaha Ātea - Space Institute, University of Auckland, Private Bag 92019, 1142, Auckland, New Zealand, email: {cristina.parigini, roberto.armellin}@auckland.ac.nz*
[(3)] *Scientific Computation Research Institute, University of La Rioja, Madre de Dios 53, 26006, Logroño, La Rioja, Spain, email: juanfelix.sanjuan@unirioja.es*

**Abstract**

This paper introduces a versatile approach for computing the risk of collision specifically tailored for scenarios featuring low relative encounter velocities, but with potential applicability across a wide range of situations. The technique employs Differential Algebra (DA) to express the non-linear dynamical flow of the initial distribution in the primary-secondary objects relative motion through high-order Taylor polynomials. The entire initial uncertainty set is subdivided into subsets through Automatic Domain Splitting (ADS) techniques to control the accuracy of the Taylor expansions. The methodology samples the initial conditions of the relative state and evaluates the polynomial expansions for each sample while retaining their temporal dependency. The classical numerical integration of the initial statistics over the set of conditions for which a collision occurs is thus reduced to an evaluation of mono-dimensional time polynomials. Specifically, samples reaching a relative distance below a critical value are identified along with the time at which this occurs. The approach is tested against a Monte Carlo (MC) simulation for various literature test cases, yielding accurate results and a consistent gain in computational time.

## 1 Introduction

With the continuous advancement of space technology, the frequency of spacecraft launches has increased steadily, leading to growing congestion in Earth's orbital environment. Over recent decades, this congestion has become increasingly hazardous to spacecraft operations, primarily due to the rising accumulation of space debris. Projections indicate that this trend will intensify, particularly with the deployment of large-scale satellite constellations, which will further saturate the most frequently used orbital regimes. Consequently, spacecraft operators are confronted with a growing number of close approach alerts and are required to perform an increasing number of Collision Avoidance Maneuvers (CAMs) to safeguard mission integrity [1].

A key aspect of managing these events is the assessment of their criticality, which mainly relies on evaluating the Probability of Collision (Pc) between the objects involved [2]. In the literature and collision avoidance practice, these are referred to as the primary (typically the controlled asset of interest) and the secondary object (the one it encounters).

The methods for computing collision risk have been tailored for different conjunction types. Close approaches between satellites are in fact typically classified as either short-term or long-term encounters [3]. Short-term encounters involve objects with significantly different orbits, resulting in encounter velocities reaching several kilometers per second near the point of closest approach. These encounters last only a few seconds at most. Throughout the encounter, the relative velocity vector remains constant in both intensity and direction, leading to a straight-line relative trajectory. Moreover, the relative velocity uncertainty is deemed negligible in comparison to its pronounced mean. Consequently, the position error combined ellipsoid remains stable throughout the encounter since the positional uncertainties of the objects can be defined by two uncorrelated constant covariance matrices. Various methods are available in the literature to compute the collision probability for the short-term case. The problem has been, in fact, extensively studied by Foster [3], Patera [5, 6], Alfriend et al. [7], Afano [8], Chan [9, 10], and more recently by Serra et al. [11].

Typically, as illustrated in Fig. 1-(a), the collision risk is calculated by integrating the Probability Density Function (PDF) of the combined positional uncertainty over the volume defined by the combined hard-body sphere [5], as it moves along the relative trajectory. Given that this trajectory is rectilinear, the swept volume can be approximated as an infinite cylinder aligned with the direction of the relative velocity. This allows the 3D integral to be reduced to a 2D one, since the integration along the direction of relative velocity yields unity. As a result, the PDF is usually evaluated on a plane perpendicular to the relative velocity vector, commonly referred to as the B-plane.

The other type of encounter occurs between two satellites traveling along neighboring orbits, such as between two GEO satellites at adjacent longitude positions or, more generally, during close satellite operations, including rendezvous, formation, and cluster flights. It is worth noting that similar geometries can also occur naturally, albeit less frequently. These close approaches are characterized by a low relative velocity, typically on the order of a few meters per second. In such scenarios, the two objects remain in close proximity for an extended duration, approximately on the order of the orbital period. As shown in Fig. 1-(b), unlike in short-term close approaches, the relative velocity vector is not constant; it evolves over time in both direction and magnitude. For that, the relative trajectory usually bends and becomes non-linear. The uncertainty in the relative state can no longer be assumed constant; it also changes over time rotating and deforming. Additionally, the uncertainty associated with the relative velocity must be accounted for and cannot be neglected anymore. This results in a time-dependent, evolving combined covariance matrix during the encounter, and the collision tube can exhibit complex, highly non-linear geometries that are very difficult to integrate.

The collision risk for long-term encounters has not been as thoroughly investigated as it has been for short-term encounters. A first category of methods, as in [12, 13] and [14], tries to solve the bending tendency of the collision tube. The key concept is to discretize the collision tube into small subsections and to consider that for each segment the assumptions of a short-term encounter still hold. A better characterization of the collision volume is outlined in the works of Chan [15, 16], where the swept-volume of the hard-body is described as an envelope of ellipsoids. However, it is important to note that these methods do not account for velocity uncertainties in the formulation of the problem. Coppola [17], on the other hand, presents a comprehensive mathematical framework that, for the first time, incorporates velocity uncertainties into the formulation. This marks one of the most extensive efforts to address the collision probability problem in a general manner. In this case, the intricate integration volume is continually mapped over time through the dynamic evolution of the initial conditions on the 3D surface of the hard-body sphere. However, the assumptions made do not accommodate for multiple encounters within the analysis timeframe, making the method not suitable for complex intersections of the integration volume.

As highlighted in [18], there is a notable absence in the literature of a general method capable of simultaneously characterizing the swept volume, especially when its shape is intricate, and computing the subsequent integral of the full relative state vector gaussian PDF over such a volume. In their work, it is suggested to approximate the swept volume using a Polynomial Superlevel Set (PSS) followed by a Monte Carlo integration to calculate the Pc.

So far, the most comprehensive methodologies capable of simultaneously accounting for uncertainties in both position and velocity, while handling complex encounter geometries, are Monte Carlo (MC) techniques [19, 20]. These methods offer a straightforward problem formulation and an intuitive approach to computing the Pc, without requiring an explicit mathematical characterization of the integration volume. Although MC simulations are often regarded as the standard for validating new Pc estimation methods due to their high accuracy, their primary drawback remains the significant computational cost. To address this, several studies have focused on speeding up the MC approach through techniques such as subset simulation [21] and line sampling [22] which use stepwise procedures based on Markov Chains to reduce the number of samples required for a given confidence level.

Rather than relying on a traditional MC approach, this work employs Differential Algebra (DA) [23] to efficiently characterize the collective behavior of subsets of sampled initial conditions. This approach builds upon the work of Morselli et al. [24], who first applied DA to estimate Pc by representing the Time of Closest Approach (TCA) and the relative distance as Taylor expansions of the initial condition distributions. Our study consolidates and extends recent developments by the authors [25], with the goal of adapting and generalizing the methodology in [24] for long-term encounter scenarios. In details, we propose to compute the dynamical evolution of the initial conditions as a patched 7D continuum, where each patch is represented by a high-order Taylor expansion in both time and the initial conditions at $t_0$. The integration accuracy of the dynamical flow is controlled by the Automatic Domain Splitting (ADS) algorithm [26], which adaptively splits the initial domain of the combined covariance to ensure precision of the Taylor approximation. Subsequently, the PDF, defined in the initial relative state space, is sampled to generate specific realizations. Each patch is evaluated accordingly, resulting in a time-dependent Taylor expansion of the miss distance. Collision probability is computed as the fraction of samples for which the miss distance falls below a predefined threshold, determined by identifying the real roots of the corresponding Taylor polynomials. The proposed methodology provides the advantages of a standard MC simulation, such as the ability to avoid restrictive assumptions, including the neglect of velocity uncertainty or the exclusion of multiple conjunctions within the screening interval. In principle, it can accommodate any form of initial uncertainty without requiring the assumption of a Gaussian distribution. Furthermore, the use of DA significantly enhances computational efficiency and offers a flexible framework well-suited for parallel implementation.

The proposed methodology forms part of the authors' broader effort to extend and enhance the capabilities of the DLR German Space Operations Center's (GSOC) Collision Avoidance System (CAS) [28], preparing it for the challenges posed by an increasingly congested orbital environment. In particular, it is designed with direct applicability to GEO assets under their control, which may at times experience long-term encounters with neighboring satellites.

The paper is structured as follows: Section 2 introduces the mathematical formulation commonly used to compute the Pc with a MC approach. Section 3 presents the core methodology, detailing how DA and ADS are employed to model the dynamical evolution of initial conditions, and how root-finding in the resulting polynomial expressions is used to determine collision events. Section 4 provides benchmark test cases from the literature, along with a comparison of computational performance against an equivalent standard MC simulation. Sub-section 4.3 applies the proposed approach to a real-world long-term conjunction scenario, offering practical guidance for operational implementation. Finally, Section 5 reports the conclusions and outlines directions for future work.

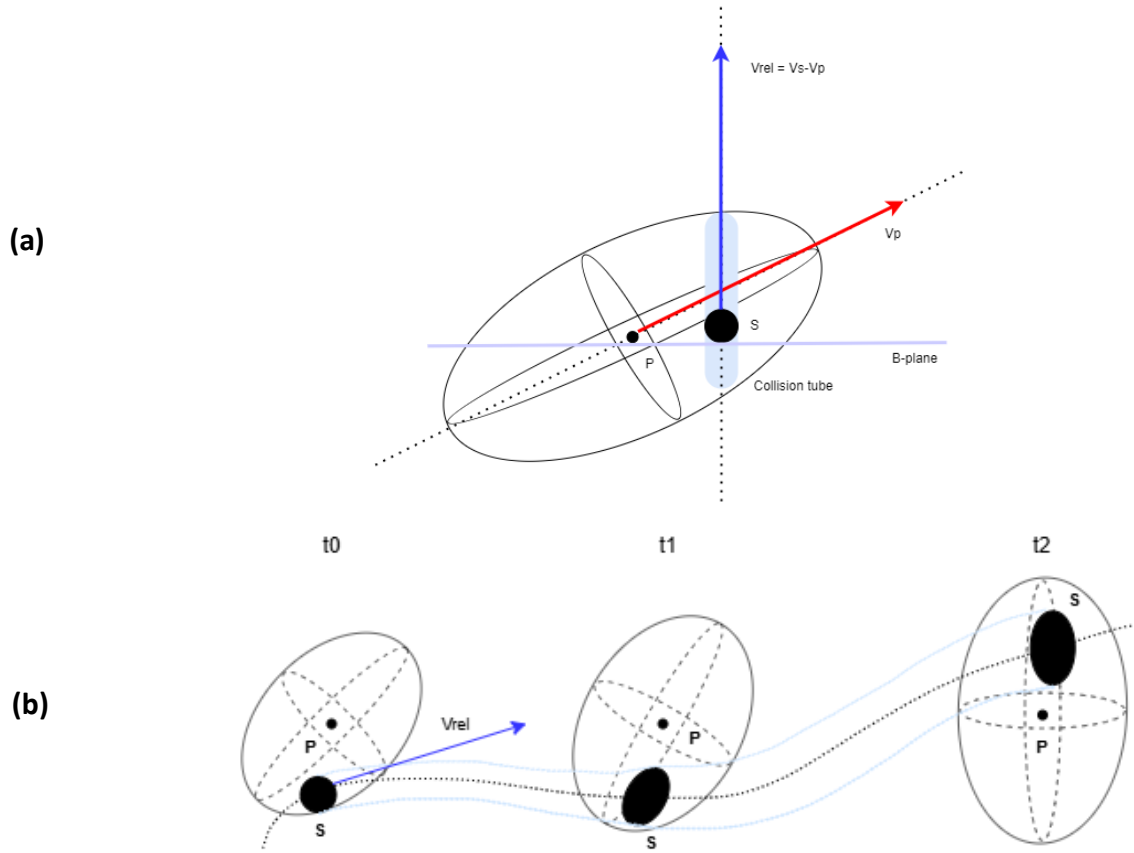

Fig. 1: representation of short-term (a) and long-term (b) encounter scenarios. By convention, the combined covariance position ellipsoid is centered on the primary object (P), while a sphere of radius equal to the Hard-Body Radius (HBR) is centered on the secondary object (S), accounting for the dimensions of both objects. In long-term encounters, the topological points on the hard-body sphere's surface at $t_0$ evolve into an ellipsoid as the combined covariance rotates and deforms over time.

## 2    Problem description

Building on the derivation in [17], the statistical event for which a collision occurs is introduced as follows: given the initial distribution of the state for two space resident objects at time $t_0$, the Hard-Body Radius $HBR$ [5], and a maximum period of interest $T$, a collision between two objects is deemed to occur if there exists a time $t$, within the interval $I = [t_0, t_0 + T]$, such that the norm of the relative distance vector $\boldsymbol{d}(t)$, is less than or equal to $HBR$. To assess the likelihood of this event occurring, and consequently characterize the Pc in a comprehensive and general manner, we introduce the relative state vector of the two objects engaged in the encounter, $x(t)$, which is a function of time. This vector is defined as

$$x(t) = \begin{pmatrix} \boldsymbol{d}(t) \\ \boldsymbol{v_{rel}}(t) \end{pmatrix}, \qquad (1)$$

in which $\boldsymbol{v_{rel}}$ is the relative velocity of the secondary object with respect to the primary. It is important to emphasize that $x(t)$ is defined as the difference, expressed in an inertial reference frame, between the individual states of the two objects involved in a close approach denoted as $x_p(t)$ and $x_s(t)$, respectively. These two quantities represent six-dimensional general multivariate random variables, which may follow arbitrary distributions. However, within the scope of this study, they are assumed to be statistically independent, implying that their cross-covariance is zero:

$$COV(x_p, x_s) = COV(x_s, x_p) = 0 . \tag{2}$$

This assumption is widely adopted in the literature and is not overly restrictive, as $x_p(t)$ and $x_s(t)$ are typically obtained through separate orbit determination processes. These processes are often independent, relying on distinct dynamical models, measurement data, and associated uncertainties.
For notation simplicity, at the initial time $t_0 = 0$, the relative state vector is denoted as $x(t_0) = x_0$. Defining the PDF of the relative state vector at $t_0$ as $\rho_0(x_0, t_0)$, we can, without loss of generality, define Pc as:

$$\text{Pc} = \Pr(x_0 \in V) = \int_V \rho_0(x_0, t_0) \, dx_0 \tag{3}$$

where $V \subseteq \mathbb{R}^6$ represents the initial set for which a collision occurs at some future time $t$. For long-term encounters, the time evolution of the set $V$ represents the same integration volume modelled in [15, 16, 18]. The set $V$ can be interpreted as a sub-region of the multi-dimensional space $\mathbb{R}^6$ wherein every realization of the random vector $x_0$ inevitably leads to the violation of the condition $\|d(t)\| \leq HBR$ at a future time $t$. In mathematics:

$$V = \{x_0 \in \mathbb{R}^6 : \exists \, t \in I : \|d(t)\| - HBR \leq 0\} \tag{4}$$

By definition of PDF, the 6D integral of the initial statistic over the set $V$ gives, in fact, the likelihood that $x_0 \in V$. Notably, the integrand in Eq. (3) represents the PDF of an arbitrary distribution. Hence, the methodology outlined in the following section is capable of handling the dynamical evolution of any statistical distribution, provided that its form is known a priori at $t_0$. Although the approach is not limited to any specific distribution, for the purposes of the present analysis, $\rho_0(x_0, t_0)$ is modelled as Gaussian. This choice is justified by the fact that, at least at $t_0$, the relative state $x_0$ typically results from an orbit determination process, for which the Gaussian assumption is generally considered reasonable. Thus:

$$\rho_0(x_0, t_0) = \frac{e^{\left(-\frac{1}{2}(x_0-\mu_0)^T P_0^{-1}(x_0-\mu_0)\right)}}{\sqrt{(2\pi)^6}\sqrt{\|P_0\|}} \tag{5}$$

where $\mu_0$ and $P_0$ represent respectively the mean and the covariance matrix of $x_0$.

One way of computing the integral in Eq. (3) is via a MC-based method. In fact, in such cases, the initial conditions $x_0$ are sampled and trajectories over the time interval $[t_0, t_0 + T]$ are computed according to some dynamical model that propagates the relative state from time $t_0$ to $T$. The dynamics are usually expressed as an Ordinary Differential Equation (ODE) of the form:

$$\begin{cases} \dot{x}(t) = f(x(t_0), x(t), u(t), t) \\ x(t_0) = x_0 \end{cases}, \tag{6}$$

where the vector $u(t)$ represents an eventually modelled maneuver in the relative dynamics.

Even when the initial conditions are modeled as Gaussian, their propagation through the dynamics described in Eq. (6), which, in the context of long-term encounters, are nonlinear, generally leads to a loss of Gaussian properties in the state distribution at future times $t$. Unlike approaches such as [17], the proposed methodology does not require the assumption that the statistical distribution remains Gaussian throughout the time interval $I$.

Finally, to compute the Pc, each sample trajectory is analyzed to verify if, at some future time, a collision occurs. If one hit is recorded for a specific sample, it means that it originally belonged to the set $V$. The probability is then computed by evaluating the ratio between the number of samples that produced a hit over the total number of samples.

## 3 Methodology

### 3.1 DA to express the dynamical evolution of $x_0$

As outlined in the preceding section, our methodology employs DA techniques to express the dynamical evolution of the initial relative conditions. DA provides a computational framework that enables the treatment of functions as $n^{th}$ order Taylor polynomial expansions within a computer environment, rather than handling them solely as floating-point values. This framework holds considerable potency as it allows to extract more information on a function rather than its mere raw values [23]. Within this context, the time $t$ can be expressed as a DA variable $\tau$ and scaled with respect to the maximum time of interest $T$, such that $\tau \in [-1,1]$:

$$\tau = \frac{2(t-t_0)}{T} - 1. \tag{7}$$

Before introducing a vector of DA variables that corresponds to the variation of the initial relative state vector from its mean at $t_0$, we perform a rotation to the space in which $x(t)$ is defined. This last may be, in general, either an inertial reference frame or a frame co-moving with the target along its trajectory. Such frames are commonly employed to express relative dynamics. Regardless of the chosen initial frame, which can be arbitrary, a transformation is applied to align the state space with the principal axes of the initial combined covariance matrix $P_0$. Specifically, the transformation is based on the eigen-decomposition of $P_0$, such that

$$P_0 = Q \Lambda Q^T, \tag{8}$$

where $Q \in \mathbb{R}^{6 \times 6}$ is an orthogonal matrix, whose columns are the eigenvectors of $P_0$, and $\Lambda \in \mathbb{R}^{6 \times 6}$ is a diagonal matrix containing the corresponding eigenvalues. This decomposition enables the mapping of a vector $\boldsymbol{m} \in \mathbb{R}^6$ from the original reference frame into an equivalent vector $\boldsymbol{m}' \in \mathbb{R}^6$, expressed in a coordinate system aligned with the principal axes of the uncertainty ellipsoid defined by $\boldsymbol{P_0}$, according to:

$$\boldsymbol{m}' = \boldsymbol{Q}^T(\boldsymbol{m} - \boldsymbol{\mu_0}). \tag{9}$$

From this point onward, all references to the relative state vector and its DA representation will be made in the transformed coordinate system. To avoid overly pedantic notation, we deliberately omit the prime symbol (').

At this point, consistently with the DA variable $\tau$, we introduce a vector of DA variables, $\delta \boldsymbol{x_0}$, representing the deviation of the initial relative state vector from $\boldsymbol{\mu_0}$. Each component is normalized by the maximum expected variation, $\Delta \boldsymbol{x_0}$, such that it is defined within the interval $[-1,1]$. This is

$$\delta \boldsymbol{x_0} = \frac{\boldsymbol{x_0} - \boldsymbol{\mu_0}}{\Delta \boldsymbol{x_0}}. \tag{9}$$

The relative state at a given scaled instant $\tau$, can now be computed via integration of the dynamics described in Eq. (6) and expressed in the DA framework as:

$$\boldsymbol{x} = \mathcal{T}_x(\tau, \delta \boldsymbol{x_0}). \tag{10}$$

That is a vector of high order polynomials that are functions of the deviations of the scaled time $\boldsymbol{\tau}$ and the initial normalized relative statistics, $\delta \boldsymbol{x_0}$. The Taylor map $\mathcal{T}_x$ establishes a relationship between the perturbed initial state vector and the corresponding state vector at a specified time within $\boldsymbol{I}$, utilizing the dynamical model defined in (6). This mapping from the initial set to the final one bears conceptual similarity to the mathematical notion of a manifold well described in [27]. Within the context of this research, the employed dynamical model may be arbitrarily complex and may include highly nonlinear relative motion. The only requirement is that it must be described by an analytical law that establishes a functional relationship between the state at time $\boldsymbol{t}$ and the initial conditions at $\boldsymbol{t_0}$.

Utilizing DA in this context offers several advantages. Firstly, it enables the representation of an infinite set solely through its Taylor expansion coefficients. This preserves a specific analytical structure in contrast to a mere point-wise set representation. Secondly, and perhaps most significantly, it allows the propagation of entire sets through a function using straightforward DA arithmetic operations. Unlike a standard MC simulation, where the ODE flow of equation (6) is integrated for each sample, in this scenario, only a single integration is required. The resulting DA expansion represents the outcome of propagating all points from the initial domain through the ODE in (6).

### 3.2 ADS to control integration accuracy

The challenging part of this approach arises from the nonlinear dynamics involved usually in long-term encounters. When the Taylor map needs to approximate a strongly non-linear function, the convergence of the ODE expansion across the domain becomes inaccurate. Consequently, the DA map, which is a local representation of the function, poorly represents the actual evolution of the whole domain, even though the description is accurate in the vicinity of the center of the expansion.

To address this issue, we employ the technique proposed in [26]. The ADS algorithm identifies instances where the ODE flow expansion over the initial conditions no longer accurately describes the dynamics. Once such a scenario is detected, the domain of the original polynomial expansion is divided along one of the expansion variables into two domains, each half the size of the original. By re-expanding the polynomials around the new center points, two separate polynomial expansions are generated.

To aid the reader's understanding and provide a visual interpretation of the concept, this process is illustrated in Fig. 2 -(a) and (b). In the visualization, a 2D function $f(x_1, x_2)$ is depicted alongside the defined domain of the variables $(x_1, x_2)$. Additionally, Taylor expansion centered around the domain's midpoint is displayed to approximate $f(x_1, x_2)$. The accuracy of the approximation is high near the center but diminishes towards the domain edges. To ensure accuracy, the algorithm iteratively splits the initial domain into two segments whenever the Taylor series representation diverges from the actual function by a user-defined margin ε. Subsequently, the expansions are recalculated around the new centers, and this process continues until all expansions accurately represent the function within the specified threshold ε.

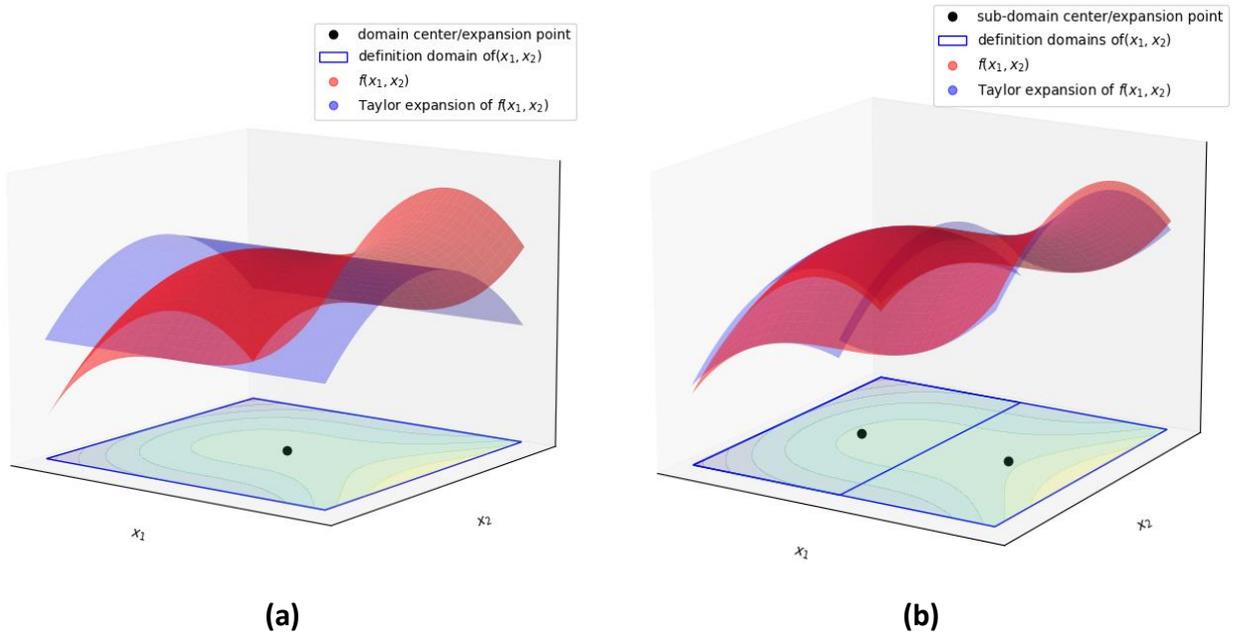

**(a)**          **(b)**

**Fig. 2: ADS algorithm illustration. (a) Taylor expansion of $f(x_1, x_2)$ around initial domain's midpoint. (b) Taylor expansions of $f(x_1, x_2)$ recalculated around the new domain centers.**

In a similar fashion, in our case the initial 7D domain, defined by the variables $\tau$ and $\delta x_0$ is split into different sub-domains. The dynamical evolution of the initial condition assumes the shape of a patched 7D continuum, mathematically defined as a manifold object [31]. Fig. 3 attempts to give a visual representation of this last, considering only the position components of the relative state vector and the time. To each patch at a given time corresponds a Taylor expansion, function of $\tau$ and of $\delta x_0$, that approximates locally the dynamical flow. Once a single integration has been performed and the evolution of the initial condition is approximated by patched polynomials, our methodology proceeds to calculate the function $d^2$ within the DA framework:

$$d^2 = \mathcal{T}_{d^2}(\tau, \boldsymbol{\delta x_0}) \tag{11}$$

Here, $\boldsymbol{d^2}$ is a high order polynomial representation of the relative distance squared expressed as function of time and initial state. It is again piece-wise defined, and its definition interval depends on the ADS accuracy control algorithm.

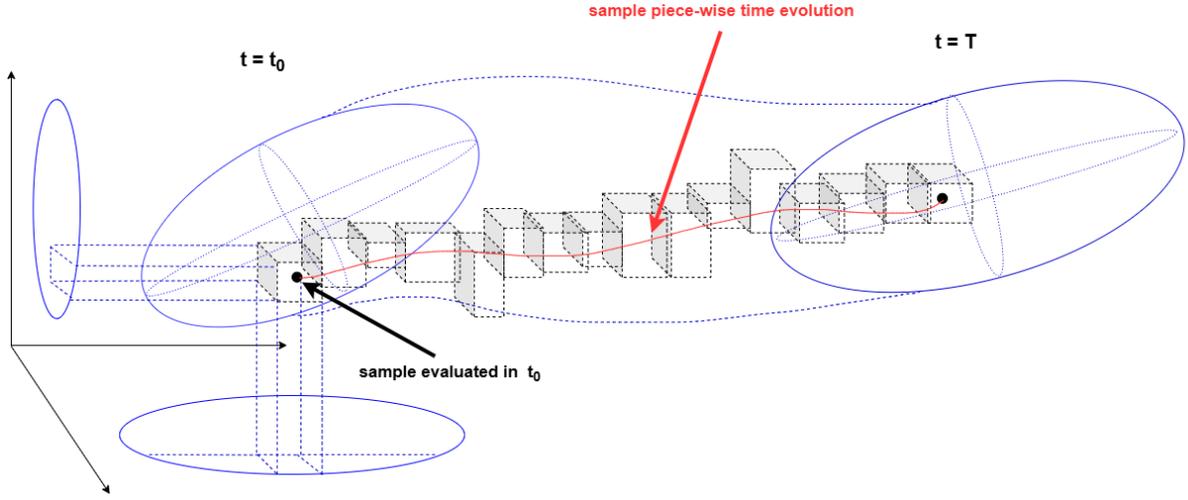

Fig. 3: Dynamical flow evolution of the initial conditions $x_0$. For representation purposes, only spatial coordinates are shown.

Subsequently, the initial set $\boldsymbol{x_0}$ is sampled. Each realization $\boldsymbol{X_{0i}}$ of the random vector is linked to its respective initial patch and evaluated only in space and velocity. This evaluation reduces the dimensions of the $d^2$ polynomials, resulting in a set of one-dimensional Taylor expansions depending solely on time:

$$d_i^2 = \mathcal{T}_{d^2}(\tau, \boldsymbol{\delta x_0} = \boldsymbol{X_{0i}}) = \mathcal{T}_{d_i^2}(\tau), \tag{12}$$

This expansion holds significant importance as it is analyzed by the algorithm to determine if a collision occurs for a specific sample. Specifically, this process is reduced to the task of locating the roots of

$$f_{d_i} = d_i^2 - HBR^2 = \mathcal{T}_{f_{d_i}}(\tau) \tag{13}$$

within the defined bounds of the Taylor expansion. This process is again illustrated in Fig. 3, where a sample is evaluated within the split initial set, resulting in a subset of one-dimensional polynomials that approximate the time evolution of the trajectory highlighted in red.

## 3.3 Finding the roots of high order one-dimensional time polynomials

At this stage, the task of determining whether a given sample trajectory results in a collision, and thus contributes to the cumulation of Pc, is reformulated as a problem of finding the roots of high-order univariate polynomials that depend solely on time. This is accomplished through the application of a series of analytical theorems, following a methodology analogous to that employed in a different context in [32]. The root-finding algorithm is detailed in Fig.4. After evaluating the sample and computing the DA expression in Eq. (13), we iterate through all the polynomials that define a sample trajectory. For each Taylor expansion and its corresponding definition interval, we initially verify if its center falls into the negative range to rule out the possibility of finding no roots due to the segment trajectory already being below the collision threshold. Subsequently, we determine the number of sign changes of the polynomial coefficients using the Budan-Fourier theorem [33]. This theorem considers the number of roots by examining the sequence of coefficient sign variations in the polynomial. Specifically, if the number of sign changes is odd, it indicates the presence of at least one real root within the polynomial's interval. In such cases, the algorithm registers a hit and proceeds to analyze the next sample. If there are no sign changes (i.e., the count is zero), we conclude that the polynomial has no real root in the interval. Thus, the algorithm proceeds to analyze the adjacent Taylor expansion as time progresses along the trajectory.

The situation differs when the number of sign changes is even. In such instances, the theorem does not provide conclusive results because the number of roots can be a multiple of an even number, potentially including zero. Therefore, the workflow further investigates using the Sturm algorithm [34]. This algorithm is a robust root isolation method that precisely determines the number of roots of a high-order polynomial within an interval by recursively performing Euclidean divisions to construct a sequence of polynomials. The sign variations in this sequence are analyzed to ascertain the number of roots. As before, if there are no roots, the algorithm proceeds to analyze the neighboring polynomials. However, if at least one root is found, a hit is recorded, and the polynomial approximation of the subsequent trajectory is studied.

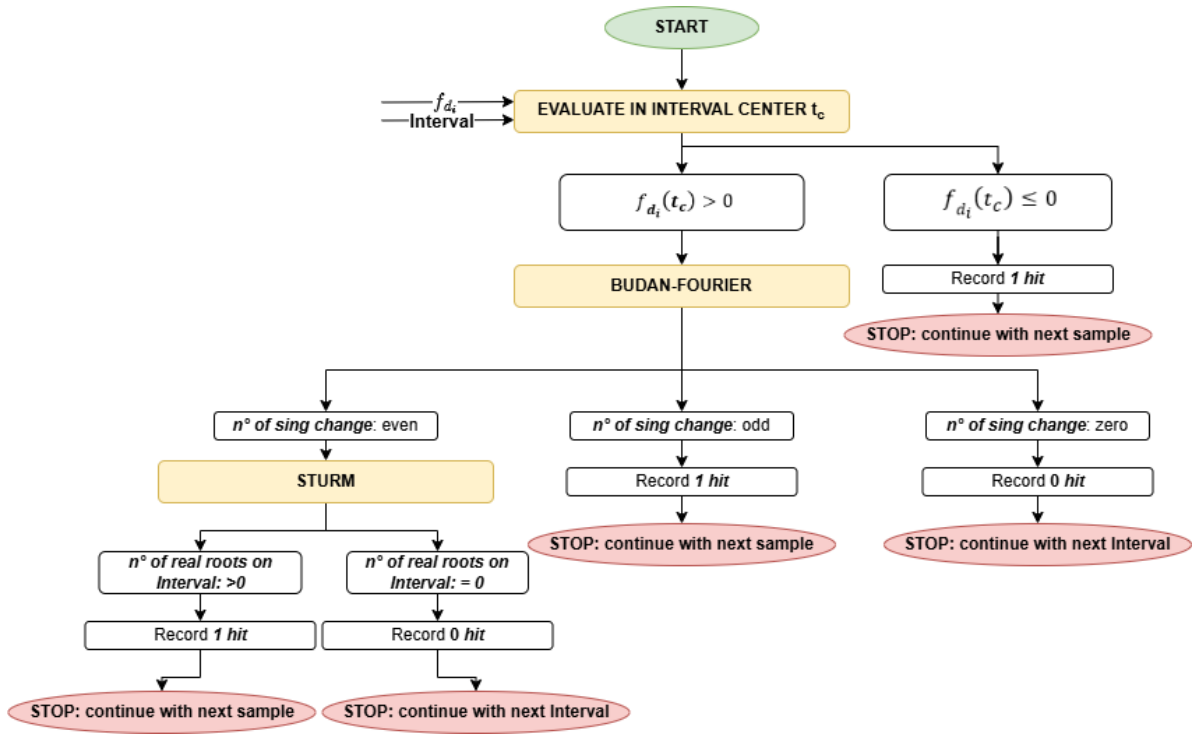

Fig.4: Roots-finding algorithm to individualize the zeros of high order mono-dimensional time polynomials

The procedure described above is executed for each sample within a loop over the ADS-defined intervals. To improve the efficiency of the algorithm, the intervals are sorted in ascending order of their initial time. Before testing whether a given interval covers the sample, the algorithm verifies that its initial time is not earlier than the final time of the previously analyzed interval; if it is, the interval is skipped without further checks.

### 3.4 Monte Carlo tool for validation

To verify our methodology and measure the efficiency of our approach, we compare the results obtained with those generated by a standard, in-house MC simulation. To this end, the initial distribution is sampled in an analogous manner and each sample is propagated from $t_0$ to $T$. A grid of equally time-spaced points in miss distance is created, followed by interpolation where the real roots of a localized cubic polynomial are extracted. To elaborate, curve fitting is executed using a technique called parabolic blending [35], where a set of four equally spaced points is utilized to construct a third-order polynomial by merging two quadratic polynomials generated from the initial three points and the last three points. The minimum of the fitted curve is then determined by extracting the roots of the polynomial's first derivative. The MC process then assesses whether a collision occurs for a given sample by checking if the relative distance at any point within the timeframe is equal to or less than the HBR. Fig. **5: High-level algorithm description for (a) DA polynomial evaluation technique and (b) standard MC tool to compute Pc** outlines the primary distinctions between our DA approach and a conventional MC-based method for computing the Pc. Essentially, our technique requires only a single integration to generate the 7D manifold and a series of polynomial evaluations for each sample. Instead, in a standard MC method, one must initially perform numerical propagation of each sample and then interpolate the grid of discrete points in relative distance.

(a)
```
Algorithm 1 DA polynomial evaluation technique
 1: DA variable initialisation.
 2: Perform ODE integration with ADS control.              ▷ eq. (10)
 3: for each sample X_{0i} i=1:N do
 4:     Evaluate f_d(τ, δx_0 = X_{0i}) within the split initial set.   ▷ eq. (13)
 5:     Get subset of polynomials that approximate sample trajectory f_{d_i}(t)
 6:     for f_{d_i}(t), I in sample trajectory do
 7:         procedure ROOT FINDING ALGORITHM(f_{d_i}(t), I)    ▷ Fig. 4
 8:             if n°real roots > 0 then
 9:                 record 1 Hit
10:             end if
11:         end procedure
12:     end for
13: end for
14: Compute P_c as: nHits/nsamples
```

**(b)**

```
Algorithm 2 Monte Carlo
    Initialize Δt and time grid.
 2: for each sample i=1:N do
        Propagate sample ∀t
 4:     Compute the relative distance ||d(t)||
        procedure PARABOLIC BLENDING INTERPOLATION(t, ||d(t)||)
 6:         if Interpolated ||d(t)|| < HBR then
                record 1 Hit
 8:         end if
        end procedure
10: end for
    Compute P_c as: nHits/nsamples
```

**Fig. 5: High-level algorithm description for (a) DA polynomial evaluation technique and (b) standard MC tool to compute Pc**

## 3.5 Parallelisation on multiple cores

The algorithm can be in principle divided in two stages. First, a single DA integration is performed to capture the complete dynamical evolution from $t_0$ to $T$. The computational cost of this step is essentially fixed, depending only on the system dynamics in (6) and the desired accuracy required by the ADS algorithm. Second, the algorithm evaluates samples in an MC-like fashion. Since each sample evolves independently, they can be distributed across CPU cores in batches. The computational time taken by the DA integration required to construct the patched manifold is negligible compared to sample evaluation and is therefore handled by the master thread. In contrast, the evaluation of sample trajectories and the root-finding procedure (Section 3.3) are parallelized across multiple CPU threads, with each thread processing a subset of samples. Finally, the hit counts from all threads are combined via an atomic operation to compute the Pc.

## 4 Testing

We examine the Pc values obtained by our approach using a set of artificial test cases from [19], with full details provided in the annex of that reference. These test cases are widely used in the literature as a benchmark for Pc computation methodologies, offering diverse scenarios suitable for both short-term and long-term analyses. Each test case provides the primary and secondary distributions at TCA in an inertial reference frame assuming a Gaussian distribution for the uncertainties in position and velocities. Given our methodology's reliance on relative dynamics, we compute the relative state and its related combined covariance at TCA and then retrieve the conditions at $t_0$.

In detail, in this research, the proposed methodology has been tested using three different analytical models of relative dynamics. The first is the Clohessy–Wiltshire model [29], which assumes a circular Keplerian orbit for the primary. The second is the Yamanaka–Ankersen model [36], which relaxes the circularity assumption on the primary's orbit and allows accurate modeling of relative motion for any eccentricity. The third is based on the Relative Orbital Elements (ROEs) theory developed by D'Amico [37, 38], which extends the Clohessy–Wiltshire formulation to include the effects of the Earth's second zonal harmonic $J_2$. All three models integrate the differential equations in Eq. (6) analytically. For the first two models, the initial relative state and combined covariance are obtained by applying the inverse of the State Transition Matrix (STM) to map the conditions at TCA back to the initial epoch $t_0$. For the third model, the procedure is conceptually similar but involves a more elaborate transformation chain. Specifically, the primary and secondary state vectors at TCA are first converted to osculating Keplerian elements, which are then transformed into mean elements using a first-order transformation that accounts for the $J_2$ perturbation. The ROEs [37] are subsequently computed and propagated backwards to $t_0$, yielding the relative state required for the analysis. The same transformation chain is applied to the covariance using a Jacobian matrix, which is computed in a computer environment through automatic differentiation. A relevant example of this approach can be found in [39], for the interested reader.

Regardless of the dynamics employed to consistently derive the conditions at $t_0$, the workflow outlined in Fig. 5-(a) is followed in all cases. First, Eq. (6) is integrated using the same dynamics adopted to generate the initial conditions, followed by the sampling of the initial statistic. The minimum number of samples required for statistical significance is then determined using the same statistical bounding criteria as presented in [22].

In the following discussion, we provide a detailed explanation of the methodology, offering insights into its operation and presenting plots corresponding to the first test cases based on Clohessy–Wiltshire dynamics. An overview of the complete set of tests conducted with different dynamical models is then summarized in Table 2.

### 4.1 Test case #1 (Clohessy–Wiltshire)

The first test case considered involves two satellites in GEO having a non-rectilinear encounter. Fig. **6**-(a) illustrates the combined positional covariance sampled at the initial time. The cyan points represent a subset of samples, denoted as $V$ in Eq. (3) and (4), for which a collision occurs within the timeframe of analysis. As depicted in Fig. 6-(b), the test case was deliberately designed so that even the mean of the initial distribution results in a collision, leading to a notably high final Pc reference value of 2.1783E-01. This elevated Pc level necessitates fewer than 16,000 Monte Carlo runs for the results to attain statistical significance.

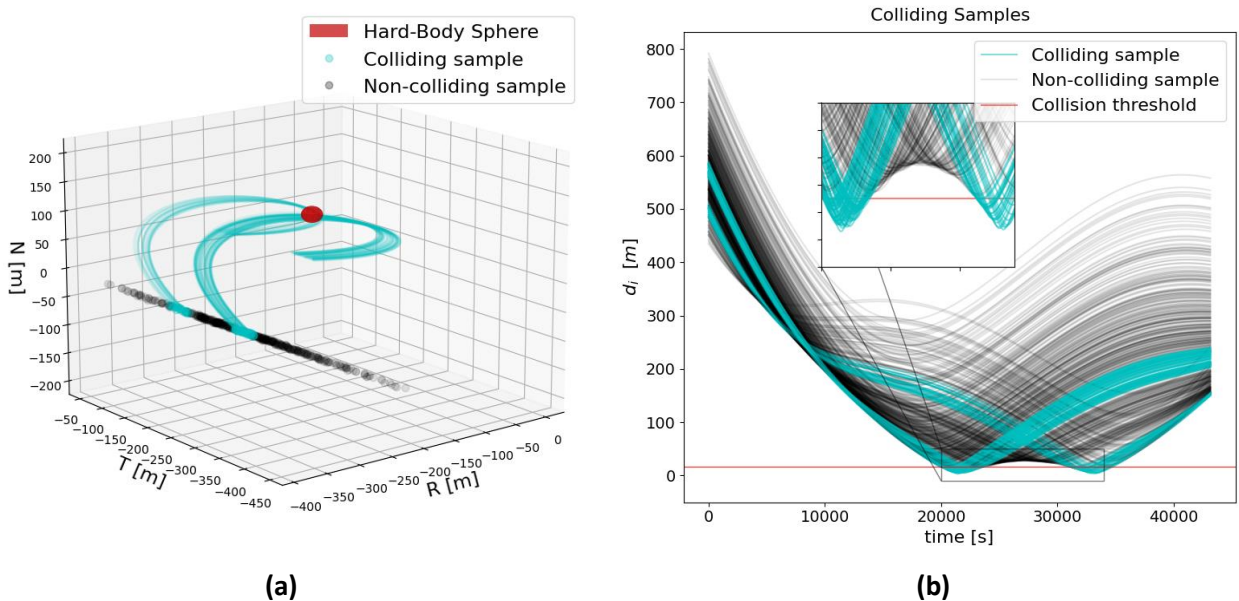

(a)          (b)

Fig. 6: (a) Samples of the combined covariance at $t_0$, with colliding samples and their trajectories shown in cyan for test case #1 (b) Evolution of the relative distance $d_i$ (in magnitude) for each sample, with trajectories that violate the collision threshold highlighted in cyan.

As described in section 3, the evolution of the initial condition generates a 7D manifold, with the initial sub-domains established by the ADS routine. Fig. 7 illustrates how the domain in the first and fifth components of the initial relative state and the time is split. Each subdomain defines the range of these variables for which a single Taylor expansion can represent the relative state vector with the required accuracy. The red line represents a particular determination of the initial relative state component, based on which all the associated sub-domains are selected.

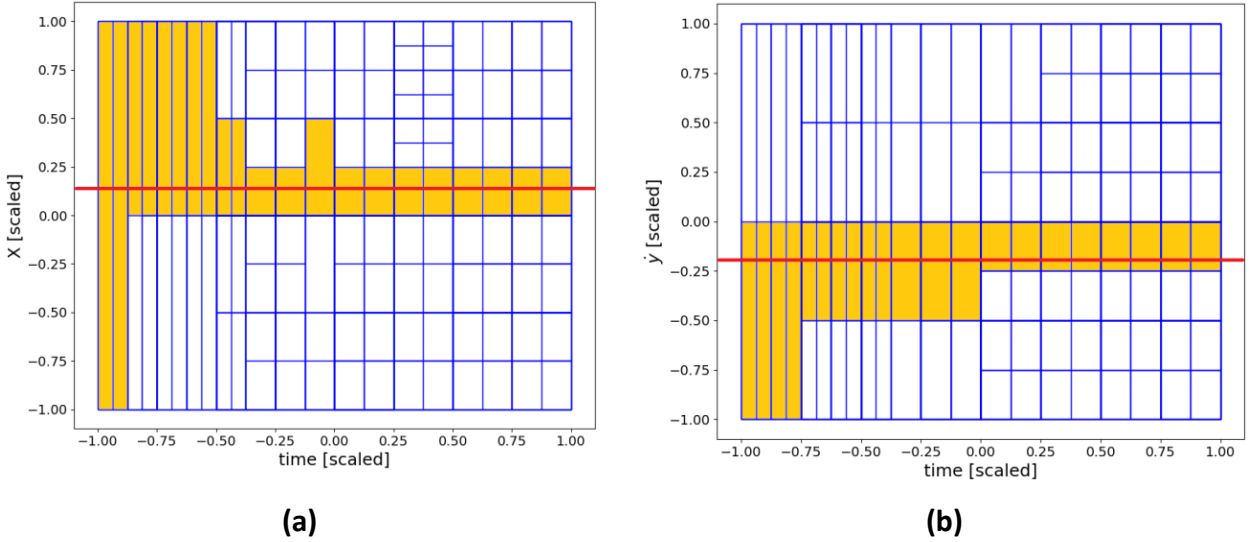

**Fig. 7: Subdomains time evolution of relative state (a) first- component and (b) fifth - component. In orange all the subdomains associated to a given sample trajectory.**

The evaluation of all these polynomials for a given sample of the initial relative state allows computing the DA expansion outlined in Eq. (13), which is a function only of the time. This is illustrated in Fig. 8-(a), where the Taylor expansions of $f_d$ are evaluated at the center of each sub-domain in time for three different samples. Two samples do not result in a collision (cyan and green curves), and one does lead to a collision (blue). The zoomed view in the figure highlights that the polynomial identifying the sample as colliding is not the one with its center below the zero line, but rather the one enclosed within the rectangle.

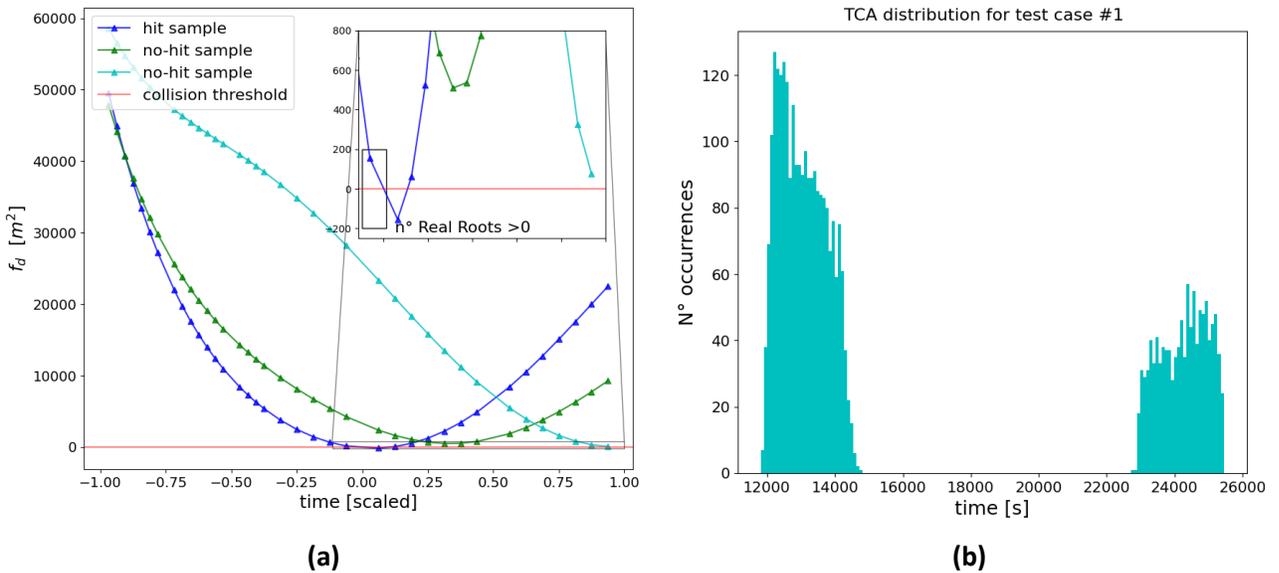

**Fig. 8: (a) Taylor expansions of $f_d$ evaluated at the centre of each time sub-domain for two "hit" samples and one "no-hit" sample. (b) TCA distribution for test case #1.**

The cumulative collision probability determined through the DA polynomial evaluation method stabilizes at a value of 2.1783e-01, mirroring the result obtained from our in-house Monte Carlo simulation. Over the analysis period, the initial conditions evolve in a manner that, on average, leads to an accumulation of Pc at two distinct times. This test case illustrates the method's ability to address multiple conjunctions and, consequently, multiple TCAs within the analysis timeframe. Fig. **8**-(b) depicts the TCA distribution in the form of a histogram, clearly indicating that among the samples resulting in a collision, a subset hits around 12,000 seconds after the start of the simulation. Successively, the Pc stops accumulating after about 2,000 seconds, and then increases again approximately 9,500 seconds later. The computational time is compared to that of the in-house Monte Carlo simulation with an integration time-step of 5 seconds. As shown in Table 2, approximately 16,000 iterations are completed in about 2.62E-01 seconds on an Intel(R) Core (TM) i9-14900K CPU @ 5.5GHz, compared to roughly 14 seconds for the MC approach, resulting in a gain in computational time of more than 5,000%.

## 4.2 Additional test cases: different dynamical models and computational time

Additional test cases were examined to assess the performance of the proposed method across a range of scenarios and with the dynamical models introduced in the section above. A brief description of each case is provided in Table 1 to give the reader an overview of its structure and purpose.

| Test case | Description |
|---|---|
| Case #2 | It shares the same characteristics and relative motion as Case #1, involving two GEO satellites; however, it features a smaller combined object radius, resulting in a mean miss distance that does not intersect the combined collision sphere. |
| Case #3 | it also considers a GEO encounter but represents a short-term conjunction, where the Pc accumulates instantaneously at TCA. |
| Case #4 | It involves non-rectilinear relative motion between two GEO objects and is designed to illustrate how the shape and orientation of the relative covariance can lead to Pc accumulation occurring after TCA. |
| Case #5 | Cases #5, #6, and #7 correspond to LEO encounters, with each case representing progressively more pronounced non-linear relative motion: from marginal non-linearity in Case #5 to strongly non-linear conditions in Case #7. |
| Case #6 | |
| Case #7 | |
| Case #8 | It features a long-term encounter in MEO. |
| Case #9 | Cases #9 and #10 involve encounters between satellites in highly eccentric orbits, with non-rectilinear relative motion and a mean miss distance at TCA that exceeds the combined object radius. Due to the high eccentricity of these orbits, only the Yamanaka-Ankersen model is used to analyze these two cases. |
| Case #10 | |

Table 1: Description of test cases.

Table 2 presents a comprehensive comparison of the DA-based methodology against a standard MC approach across various dynamical models. For each test case, the table reports the number of samples used, the final Pc obtained, the percentage error in Pc relative to the MC result, the respective computational times, and the time required by the DA and ADS components to integrate Eq. (6). Test case #7 is included in the description above (Table 1) for completeness, but it is not further analyzed due to the elevated number of samples required and its equivalence to cases #5 and #6. Moreover, cases #9 and #10 are only applicable to the Yamanaka-Ankersen dynamics, since in both scenarios the primary follows a highly eccentric orbit.

The results show that the DA-based methodology achieves excellent agreement with the MC simulation across all dynamical regimes, consistently reproducing the Pc with high accuracy. In some cases, there is a negligible discrepancy in the final Pc with an error that never exceeds 0.0176%. These minor deviations are attributed to an inherent limitation of the DA framework: it requires setting a threshold for the uncertainty in the initial deviation $\delta x_0$ to construct the patched 7D manifold. In this study, the threshold has been conservatively set to five standard deviations of the initial uncertainty. In rare cases, especially when a large number of samples is involved, a few may fall outside this 5σ boundary at $t_0$. Such samples are discarded and not included in the analysis, which can lead to a slight under-/ or over-estimation of Pc compared to the complete MC reference. Nevertheless, the overall accuracy and efficiency of the DA approach remain robust across all test scenarios.

In case #3, across all dynamics, the error with respect to the reference never exceeds 0.5%, as the test case essentially represents a 2D encounter. In such situations, the Pc accumulates almost instantaneously at TCA, requiring the algorithm to perform a large number of splits around that time to properly capture the dynamical evolution, with a corresponding reduction in the splitting tolerance ε. This leads to a loss in computational efficiency, highlighting the inevitable trade-off between runtime and Pc estimation accuracy in 2D cases. However, this limitation is not particularly critical, since for short-term encounters the use of a MC based approach is generally not recommended, since far more efficient methods exist to compute Pc with high accuracy.

In terms of computational speed, it is important to note that the DA integration consistently requires significantly less time, often by several orders of magnitude, than the evaluation of the individual samples, to the point where its contribution becomes practically negligible. This highlights a fundamental strength of the methodology: it captures the full evolution of the initial condition set through a single, efficient integration. For a visual comparison of computational times between the DA-based approach and the MC method, the reader is referred to Fig. 9. All computational times were obtained using the same hardware configuration described in the previous section.

The plot displays computational time on the left axis (in logarithmic scale) and the percentage gain in computational time relative to the corresponding MC simulation on the right axis. Overall, the DA-based approach outperforms the MC method across all test cases and dynamical models, with computational time gains ranging from a minimum of approximately 150% to nearly 60,000%. Notably, the gap in computational efficiency between the DA-based method and the MC approach widens as the complexity of the underlying dynamics increases. While the MC method may benefit from slightly faster sample propagation in simpler models, such as Clohessy–Wiltshire or the D'Amico formulation, the DA-based approach becomes significantly more advantageous as the dynamics involve more complex transformations, as in the case of Yamanaka-Ankersen.

| Clohessy-Wiltshire | | | | | | |
|---|---|---|---|---|---|---|
| Case | n° Samples | DA Polynomials | | | Monte Carlo | | Pc err [%] |
| | | Pc | Cmp. Time [s] | ADS integration time [s] | Pc | Cmp. Time [s] | |
| # 1 | 1.57E+04 | 2.1783439490e-01 | 2.2803e-01s | 3.699e-03 | 2.1783439490e-01 | 1.3612e+01 | 0.0000 |
| # 2 | 6.52E+06 | 1.6044057777e-02 | 1.4247e+02 | 7.310e-03 | 1.6043137529e-02 | 5.2175e+03 | 0.0057 |
| # 3 | 9.54E+05 | 1.0017956031e-01 | 4.4498e+01 | 1.3565e-02 | 9.9781236209e-02 | 7.2086e+02 | 0.3991 |
| # 4 | 1.24E+06 | 7.4077658315e-02 | 1.6199e+01 | 3.1722e-03 | 7.4077658315e-02 | 1.1144e+03 | 0.0000 |
| # 5 | 2.30E+06 | 4.9284041611e-02 | 2.3036e+02 | 3.8621e-02 | 4.9277954638e-02 | 2.1331e+03 | 0.0123 |
| # 6 | 2.45E+07 | 3.2113120024e-02 | 4.4939e+03 | 7.7043e-02 | 3.2113120024e-02 | 2.3475e+04 | 0.0000 |
| # 8 | 3.11E+06 | 3.5209770838e-02 | 9.8045e+01 | 7.0816e-03 | 3.5209770838e-02 | 3.0456e+03 | 0.0000 |
| Yamanaka-Ankersen | | | | | | |
| # 1 | 1.57E+04 | 2.1974522293e-02 | 7.7953e-01 | 7.8263e-02 | 2.1974522293e-02 | 2.4096e+02 | 0.0000 |
| # 2 | 6.52E+06 | 1.5625208494e-02 | 1.5904e+02 | 3.1008e-02 | 1.5623828121e-02 | 9.4998e+04 | 0.0088 |
| # 3 | 9.54E+05 | 1.0004958087e-01 | 7.8968e+01 | 4.9884e-02 | 9.9618761904e-02 | 1.3478e+04 | 0.4325 |
| # 4 | 1.24E+06 | 7.3597818057e-02 | 1.9757e+01 | 4.5257e-02 | 7.3597818057e-02 | 1.8311e+04 | 0.0000 |
| # 5 | 2.30E+06 | 4.9344170491e-02 | 3.5407e+02 | 1.6887e-01 | 4.9335474794e-02 | 3.5586e+04 | 0.0176 |
| # 6 | 2.45E+07 | 3.2108784306e-02 | 5.1490e+03 | 3.0087e-01 | 3.2108784306e-02 | 4.4743e+05 | 0.0000 |
| # 8 | 3.11E+06 | 3.5504638603e-02 | 8.7062e+01 | 4.8339e-02 | 3.5504638603e-02 | 4.6758e+04 | 0.0000 |
| # 9 | 1.80E+05 | 3.6322070245e-01 | 1.5878e+01 | 2.2812e-01 | 3.6322070245e-01 | 2.5490e+03 | 0.0000 |
| # 10 | 1.92E+05 | 3.6289961354e-01 | 1.6699e+01 | 2.1105e-01 | 3.6289961354e-01 | 2.6738e+03 | 0.0000 |
| **D'Amico ROEs (J2 only)** | | | | | | | |
| # 1 | 1.57E+04 | 2.1229299363e-01 | 1.1358e+01 | 2.9052e-01 | 2.1229299363e-01 | 4.0226e+01 | 0.0000 |
| # 2 | 6.52E+06 | 1.5595278886e-02 | 5.0212e+03 | 2.9113e-01 | 1.5595278886e-02 | 1.6352e+04 | 0.0000 |
| # 3 | 9.54E+05 | 9.9111422782e-02 | 7.9797e+02 | 4.8696e-01 | 9.8611421210e-02 | 2.6991e+03 | 0.5070 |
| # 4 | 1.24E+06 | 7.3588081261e-02 | 1.0374e+03 | 2.8894e-01 | 7.3588081261e-02 | 3.1196e+03 | 0.0000 |
| # 5 | 2.30E+06 | 1.3047958428e-01 | 1.8123e+03 | 4.4447e+00 | 1.3046827989e-01 | 6.8514e+03 | 0.0087 |
| # 6 | 2.45E+07 | 3.214569346e-02 | 4.4890e+04 | 4.9542e+00 | 3.214569346e-02 | 2.6569e+05 | 0.0000 |
| # 8 | 3.11E+06 | 3.5636115989e-02 | 5.3679e+03 | 8.5806e-01 | 3.5635151356e-02 | 1.2422e+04 | 0.0027 |

**Table 2: Test cases overview for different scenarios and different relative dynamical models.**

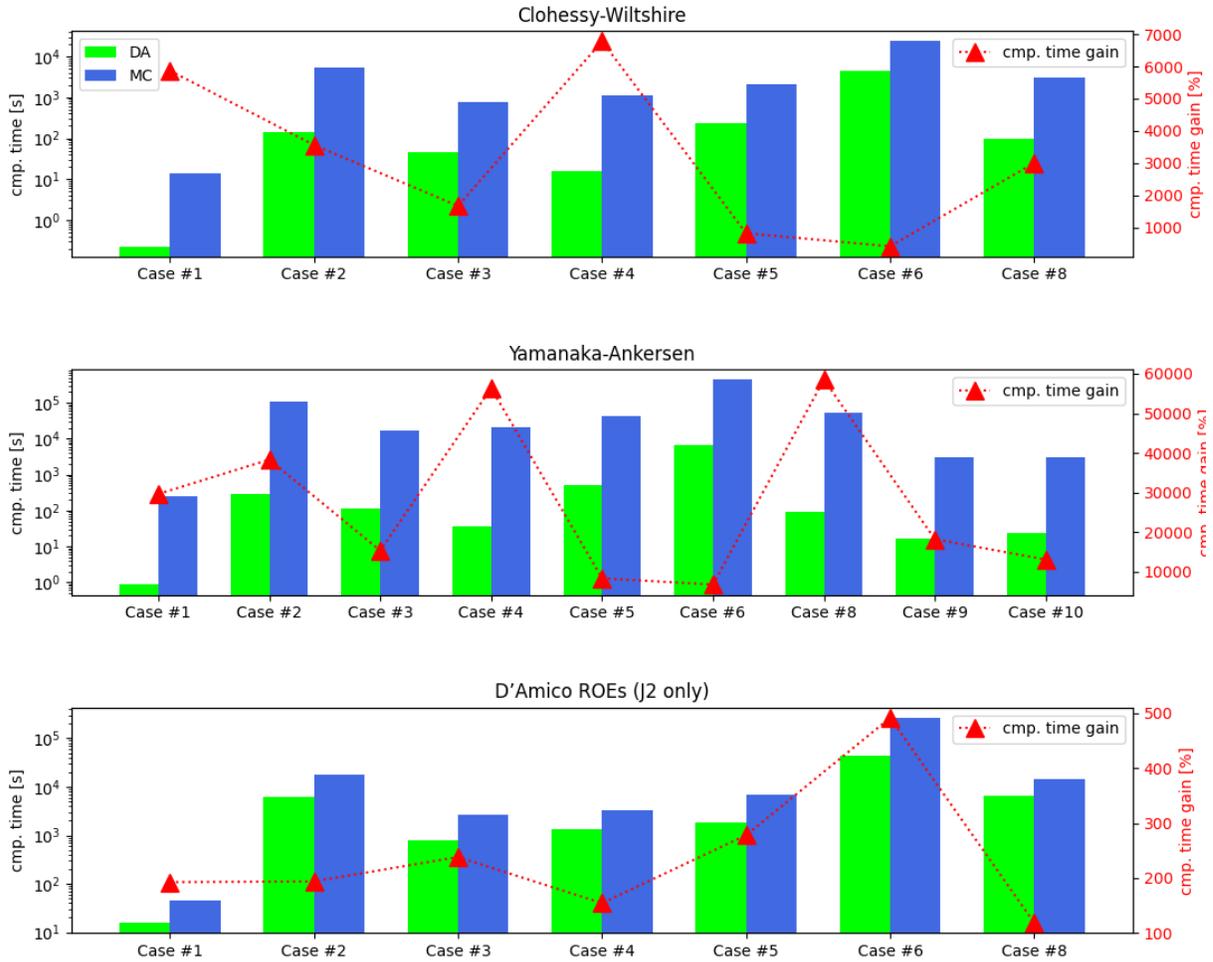

**Fig. 9:** Comparison of computational times for different test cases between the DA-based methodology and the corresponding MC simulation. The left axis shows the computational time in logarithmic scale, while the right axis shows the percentage gain in computational time.

### 4.3 Real test case

We now consider a real conjunction scenario, as the test cases presented in reference [19], while academically relevant, are not operationally realistic due to their extremely high Pc. The event involves a close approach between an asset in GEO and a secondary object in a very similar orbit, with a relative distance of 1.83 km and an HBR of 27.05 m. The encounter features a normal component of just 81 meters and a relative velocity at TCA of approximately 3 m/s, classifying it as a long-term conjunction.

For this event, the TCA conditions are summarized in Table 3. Specifically, the state vectors are provided in an ECI frame, while the corresponding covariance matrices are expressed in the RTN frame for each object. Fig. 10-(a) illustrates the evolution of the relative distance between the two objects over a 2-day period centered on the TCA. This interval reveals multiple close approaches, indicated by local minima in the relative distance. In addition to the situation at TCA, two notable minima (shown in the plot as local minimum A and B) occur approximately 24 hours before and after the closest approach. In the first one, the objects reach a separation of 10 km with a normal component of 178 meters; in the second, the normal separation is 221 meters. The instantaneous Pc, computed using the 2D methodology reported in [2], peaks at 1.42E-04 at TCA, while at the other minima, it is roughly an order of magnitude lower.

The methodology proposed in this work is tested on this scenario, and the results are compared to a MC simulation as done for the other test cases. Based on the Instantaneous Pc at TCA, the required number of samples has been estimated: approximately 3.0E+07 are needed to achieve statistical relevance with 5% accuracy and 95% confidence. The resulting Pc evolution is given in Fig. 10-(b) and (c). As it can be seen the Pc accumulates first around the first minimum, reaching the value of 1.76E-06, then accumulates for a period of 20 minutes, bracketing TCA and then accumulates again at the third minimum, settling to a value of 1.4173E-04. All the samples that are producing a hit in the MC simulation are producing a hit also in the DA methodology reaching a 0.0% estimation error with respect to the MC.

| Real long-term conjunction scenario ||
|---|---|
| State vector and covariance at epoch time ||
| Primary object | Secondary object |
| X = 35037.184606 [km] | X = -35035.891661 [km] |
| Y = -23466.448768 [km] | Y = -23467.794741 [km] |
| Z = 15.784810 [km] | Z = 15.864584 [km] |
| X_DOT = 1.710187 [km/s] | X_DOT = 1.709784 [km/s] |
| Y_DOT = -2.554690 [km/s] | Y_DOT = -2.554927 [km/s] |
| Z_DOT = -0.001790 [km/s] | Z_DOT = 0.000755 [km/s] |
| CR_R = 1.341040E+03 [m**2] | CR_R = 1.353645E+05 [m**2] |
| CT_R = -1.285933E+02 [m**2] | CT_R = -3.104145E+05 [m**2] |
| CT_T = 2.221449E+04 [m**2] | CT_T = 2.118139E+07 [m**2] |
| CN_R = 1.062929E+04 [m**2] | CN_R = 5.530171E+02 [m**2] |
| CN_T = 2.613577E+02 [m**2] | CN_T = -2.019151E+03 [m**2] |
| CN_N = 8.435330E+04 [m**2] | CN_N = 4.600785E+04 [m**2] |
| CRDOT_R = 0.000000E+00 [m**2/s] | CRDOT_R = 1.620946E+01 [m**2/s] |
| CRDOT_T = 0.000000E+00 [m**2/s] | CRDOT_T = -1.582021E+03 [m**2/s] |
| CRDOT_N = 0.000000E+00 [m**2/s] | CRDOT_N = 1.073235E-01 [m**2/s] |
| CRDOT_RDOT = 0.000000E+00 [m**2/s**2] | CRDOT_RDOT =1.189744E-01 [m**2/s**2] |
| CTDOT_R = 0.000000E+00 [m**2/s] | CTDOT_R = -9.649515E+00 [m**2/s] |
| CTDOT_T = 0.000000E+00 [m**2/s] | CTDOT_T = 1.190034E+01 [m**2/s] |
| CTDOT_N = 0.000000E+00 [m**2/s] | CTDOT_N = -3.914278E-02 [m**2/s] |
| CTDOT_RDOT = 0.000000E+00 [m**2/s**2] | CTDOT_RDOT = -3.817790E-04 [m**2/s**2] |
| CTDOT_TDOT = 0.000000E+00 [m**2/s**2] | CTDOT_TDOT = 6.930053E-04 [m**2/s**2] |
| CNDOT_R = 0.000000E+00 [m**2/s] | CNDOT_R = -2.373264E-02 [m**2/s] |
| CNDOT_T = 0.000000E+00 [m**2/s] | CNDOT_T = -4.966605E-01 [m**2/s] |
| CNDOT_N = 0.000000E+00 [m**2/s] | CNDOT_N = -3.501316E-01 [m**2/s] |
| CNDOT_RDOT = 0.000000E+00 [m**2/s**2] | CNDOT_RDOT = 4.222343E-05 [m**2/s**2] |
| CNDOT_TDOT = 0.000000E+00 [m**2/s**2] | CNDOT_TDOT = 1.972771E-06 [m**2/s**2] |
| CNDOT_NDOT = 0.000000E+00 [m**2/s**2] | CNDOT_NDOT = 9.341607E-05 [m**2/s**2] |

**Table 3: State vectors and associated uncertainties of real test case scenario.**

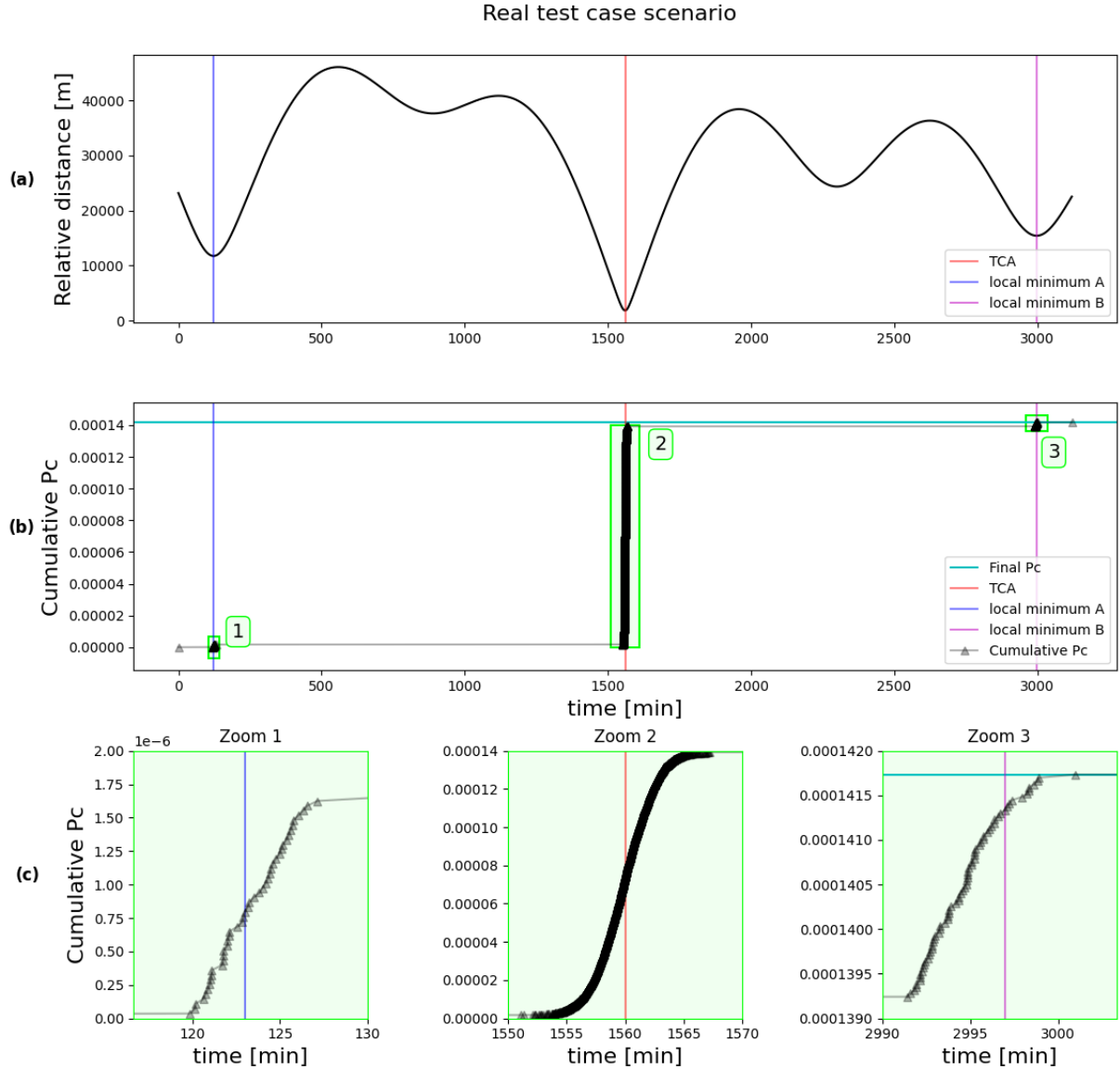

Fig. 10: (a) Evolution of relative distance for real test case. (b) Evolution of Pc. (c) Zoomed view highlighting regions of Pc accumulation.

## 5   Conclusion and future work

We presented a general methodology for computing the Pc between two space objects, particularly suited for long-term encounters with relative velocities on the order of meters per second, yet in principle applicable to two-dimensional scenarios as well. The approach employs DA to capture the non-linear time evolution of the multivariate initial relative state vector. This last is modelled as a patched 7D continuum, where each patch is represented by a high-order Taylor expansion in both time and the initial conditions at $t_0$. The integration accuracy of the dynamical flow is controlled by the ADS algorithm, which adaptively splits the initial domain of the combined covariance to ensure precision of the Taylor approximation. The initial PDF is then sampled and each patch is evaluated in a specific state-vector realization to definitively determine the time evolution of a given sample. The identification of collisions is reduced to finding the real roots of the DA approximation of the miss distance.

The proposed methodology retains all the advantages of a MC simulation while offering a substantial improvement in computational speed, owing to its DA foundation. In principle, it can be applied to any form of initial uncertainty, provided that the initial PDF is known a priori, without requiring the restrictive assumption of Gaussianity at $t_0$ or throughout the evolution of the initial conditions. Moreover, the approach can incorporate uncertainties in velocity and naturally accounts for multiple conjunctions occurring within the screening interval. Additionally, while this study has focused on three specific dynamical models, the technique is applicable to any form of relative or absolute dynamics, provided that an analytical formulation of the dynamics is available.

The technique has been extensively validated against benchmark test cases from the literature. In all scenarios, the estimated Pc shows excellent agreement with the corresponding MC simulations used for validation. The computational time gains are consistently significant, ranging from approximately 150% to nearly 60,000%, with the gap in efficiency between the DA-based method and the MC approach increasing as the complexity of the underlying dynamics grows.

This work also presents a real-case scenario involving an encounter between two GEO asset involved in a long-term conjunction. The proposed methodology successfully captures multiple conjunction events and the corresponding Pc accumulation occurring in different intervals within the computation window.

In future work, we plan to enhance the fidelity of the dynamics by extending the presented models to include additional orbital perturbations, particularly those relevant to the GEO environment.

While the reduction in computational time is evident, a key objective remains the adaptation of the methodology for operational use at GSOC. Having successfully passed an extensive validation phase, the next goal is its full integration into the CAS, enabling the handling of long-term conjunctions across multiple missions simultaneously. In doing so, the system will be capable of providing outputs that function as reliable decision-support tools for flight dynamics engineers in real-world encounter scenarios, ultimately contributing to the safe operation of more than 40 satellites.

## 6    Disclosure statement

Artificial intelligence tools have been utilized to enhance the grammar, spelling and overall readability of this manuscript.

## 7    Acknowledgments

The authors would like to thank their colleagues from the On-Orbit Servicing group for their support and the valuable discussions on the relative dynamics models presented in this work. Special thanks go to Adam Evans and the team from the University of Auckland for their assistance in parallelizing the DA-based computation of Pc, and to Thomas Caleb for his support with the root-finding algorithm. The University of La Rioja would also like to acknowledge the support of Project PID2021-123219OB-I00, funded by MICIU/AEI/10.13039/ 501100011033 and by ERDF/EU.

## 8    References

1. ESOC (European Space Operations Centre), "ESA's annual space environment report", ESA Space Debris Office (2025), ESA TR GEN-DBLOG-00288-OPS-SD

2. Klinkrad, H., "Space Debris – Models and Risk Analysis". Springer, 2006, Berlin Heidelberg.

3. Chan, K., "Short-term vs. long-term spacecraft encounters.", AIAA/AAS Astrodynamics Specialist Conference and Exhibit, 2004, Providence, RI, Paper AIAA-2004-5460

4. Foster, J. A., Estes, H. S. "Parametric analysis of orbital debris collision probability and maneuver rate for space vehicles." Technical Report. NASA JSC, 1992.

5. Patera, R. P. "General method for calculating satellite collision probability." Journal of Guidance, Control, and Dynamics, 2001, 24(4): 716–722.

6. Patera, R. P. "Calculating collision probability for arbitrary space vehicle shapes via numerical quadrature." Journal of Guidance, Control, and Dynamics, 2005, 28(6): 1326–1328.

7. Alfriend, K. T., Akella, M. R., Frisbee, J., Foster, J. L., Deok-Jin, L., Wilkins, M. "Probability of collision error analysis." Space Debris, 1999, 1(1): 21–35.

8. Alfano, S. "A numerical implementation of spherical object collision probability." The Journal of the Astronautical Sciences, 2005, 53(1): 103–109.

9. Chan, F. K. Spacecraft Collision Probability. Reston, VA, USA: Aerospace Press, 2008.

10. Chan, K.,."Collision probability analyses for earth orbiting satellites". Adv. Astronaut. Sci. 96, 1033–1048, 1997

11. Serra, R., Arzelier, D., Joldes, M., Lasserre, J. B., Rondepierre, A., Salvy, B. "Fast and accurate computation of orbital collision probability for short-term encounters." Journal of Guidance, Control, and Dynamics, 2016, 39(5): 1009–1021

12. Patera, R. P. "Satellite collision probability for nonlinear relative motion." Journal of Guidance, Control, and Dynamics, 2003, 26(5): 728–733.

13. Patera, R. P. "Collision probability for larger bodies having nonlinear relative motion." Journal of Guidance, Control, and Dynamics, 2006, 29(6): 1468–1472.

14. Alfano, S. "Addressing nonlinear relative motion for spacecraft collision probability." In: Proceedings of the AIAA/AAS Astrodynamics Specialist Conference and Exhibit, 2006: 6760

15. F.K. Chan. "Hovering collision probability." In AAS/AIAA Space Flight Mechanics Meeting, number AAS 15-234, Williamsburg, VA, USA, January 2015.

16. K. Chan. "Spacecraft collision probability for long-term encounters." Number AAS 03-549, Big Sky, Montana, USA, 2003.

17. V.T. Coppola. "Including Velocity Uncertainty in the Probability of Collision between Space Objects." Advances in the Astronautical Sciences, 143, 2012.

18. Denis Arzelier, Florent Bréhard, Mioara Joldeş, Jean-Bernard Lasserre, Sohie Laurens, et al. „Poly nomial superlevel set approximation of swept-volume for computing collision probability in space encounters." 2021. hal-03158347

19. Alfano S., Satellite conjunction Monte Carlo analysis. Advances in the Astronautical Sciences, 134:2007–2024, 2009.

20. Hall, D.T., Casali, S.J., Johnson, L.C., Skrehart, B.B., & Baars, L.G., "High Fidelity Collision Probabilities Estimated Using Brute Force Monte Carlo Simulations", 2018, https://api.semanticscholar.org/CorpusID:126009860

21. Au, S.-K., Beck, J.L.,. "Estimation of small failure probabilities in high dimensions by subset simulation." Probab. Eng. Mech. 16 (4), 263–277, 2001.

22. Koutsourelakis P.S., Pradlwarter H.J., Schuëller G.I.,"Reliability of structures in high dimensions, part I: algorithms and applications", Probabilistic Engineering Mechanics, Volume 19, Issue 4, 2004, Pages

409-417, ISSN 0266-8920, https://doi.org/10.1016/j.probengmech.2004.05.001.

23. Armellin, R., Di Lizia, P., Bernelli Zazzera, F., Berz, M. "Asteroid close encounters characterization using differential algebra: the case of Apophis." Celest. Mech. Dyn. Astron. 107, 451–470 (2010)

24. Morselli A., Armellin R., Di Lizia P., Bernelli Zazzera F., "A high order method for orbital conjunctions analysis: Monte Carlo collision probability computation", Advances in Space Research, Volume 55, Issue 1, 2015, Pages 311-333, ISSN 0273-1177, https://doi.org/10.1016/j.asr.2014.09.003.

25. A. Zollo, C. Parigini, R. Armellin, J. F. San Juan Díaz, S. Aida, R. Kahle, "Long-term collision probability computation through high order polynomials evaluation", 29th International Symposium on Space Flight Dynamics (ISSFD), Darmstadt, Germany, 22-26th April 2024

26. A. Wittig, P. Di Lizia, R. Armellin, K. Makino, F. Bernelli-Zazzera, and M. Berz, "Propagation of large uncertainty sets in orbital dynamics by automatic domain splitting," Celestial Mechanics and Dynamical Astronomy, Vol. 122, No. 3, 2015, pp. 239–261.

27. Whitney, H., "Differentiable Manifolds." Annals of Mathematics, vol. 37, no. 3, 1936, pp. 645–80., https://doi.org/10.2307/1968482.

28. Aida S., "Conjunction risk assessment and avoidance maneuver planning tools.", 6th International Conference on Astrodynamics Tools and Techniques, Darmstadt, Germany, March 2016.

29. Clohessy, W. H.; Wiltshire, R. S. (1960). Terminal Guidance System for Satellite Rendezvous. Journal of the Aerospace Sciences. 27 (9): 653–658. doi:10.2514/8.8704

30. Alfano, S. "Determining Satellite Close Approaches," Journal of the Astronautical Sciences, Vol. 41, No. 2, April-June 1993, pp. 217-225.

31. Whitney, H., "Differentiable Manifolds." Annals of Mathematics, vol. 37, no. 3, 1936, pp. 645–80., https://doi.org/10.2307/1968482.

32. Caleb, T., Losacco, M., Fossà, A. et al.," Differential algebra methods applied to continuous abacus generation and bifurcation detection: application to periodic families of the Earth–Moon system". Nonlinear Dynamics, Volume 111, pp. 9721–9740, 2023, https://doi.org/10.1007/s11071-023-08375-0

33. Fourier, Jean Baptiste Joseph (1820). "Sur l'usage du théorème de Descartes dans la recherche des limites des racines". Bulletin des Sciences, par la Société Philomatique de Paris: 156–165

34. Sturm, Jacques Charles François (1829). "Mémoire sur la résolution des équations numériques". Bulletin des Sciences de Férussac. 11: 419–425.

35. Alfano, S. and Gandhi, A. M., "Fitting atmospheric parameters using parabolic blending", Journal of Aircraft, 1990, https://doi.org/10.2514/3.45986Alfano, S.

36. Yamanaka, K. and Ankersen, F.,"New State Transition Matrix for Relative Motion on an Arbitrary Elliptical Orbit", Journal of Guidance, Control, and Dynamics, Volume 25, pp. 60-66, 2002, https://doi.org/10.2514/2.4875

37. D'Amico, S.," Autonomous formation flying in low earth orbit." Ridderkerk: Ridderprint bv, 2010. 209 p.

38. Sullivan, J., Grimberg, S. and D'Amico, S., "Comprehensive Survey and Assessment of Spacecraft Relative Motion Dynamics Models", Journal of Guidance, Control, and Dynamics, Volume 40, 2017, https://doi.org/10.2514/1.G002309

39. Acciarini, G., Atılım, G. B., Izzo, D., "Closing the gap between SGP4 and high-precision propagation via differentiable programming", Acta Astronautica, Volume 226, Part 1, 2025, Pages 694-701, ISSN 0094-5765, https://doi.org/10.1016/j.actaastro.2024.10.063.